%% file: template.tex
\documentclass[a4paper]{article}

\usepackage{INTERSPEECH2020}
\usepackage[acronym,nomain,nonumberlist]{glossaries}
\usepackage{hyperref}
\input{glossaries}
\graphicspath{{figures/}}
\usepackage{float}
\usepackage{stfloats}
\usepackage{multirow}
\usepackage{rotating}
\usepackage{booktabs}
\usepackage{url}
\title{Self-Attention Encoding and Pooling for Speaker Recognition}
\name{Pooyan Safari, Miquel India, Javier Hernando}
\address{TALP Research Center\\ Universitat Politecnica de Catalunya, Barcelona, Spain}
\email{pooyan.safari@tsc.upc.edu,\{miquel.angel.india,javier.hernando\}@upc.edu}

\begin{document}

\maketitle
\begin{abstract}
%   While mobile-based end-user applications become more and more popular, 
  The computing power of mobile devices limits the end-user applications in terms of storage size, processing, memory and energy consumption. 
%   On the contrary, the Deep Learning approaches usually are energy expensive, demanding storage, processing power, and memory. 
  These limitations motivate researchers for the design of more efficient deep models. On the other hand, self-attention networks based on \textit{Transformer} architecture have attracted remarkable interests due to their high parallelization capabilities and strong performance on a variety of \gls{nlp} applications. Inspired by the \textit{Transformer}, we propose a tandem Self-Attention Encoding and Pooling (SAEP) mechanism to obtain a discriminative speaker embedding given non-fixed length speech utterances. SAEP is a stack of identical blocks solely relied on self-attention and position-wise feed-forward networks to create vector representation of speakers. This approach encodes short-term speaker spectral features into speaker embeddings to be used in text-independent speaker verification. We have evaluated this approach on both \textit{VoxCeleb1 \& 2} datasets. The proposed architecture is able to outperform the baseline x-vector, and shows competitive performance to some other benchmarks based on convolutions, with a significant reduction in model size. It employs $94\%$, $95\%$, and $73\%$ less parameters compared to ResNet-34, ResNet-50, and x-vector, respectively. This indicates that the proposed fully attention based architecture is more efficient in extracting time-invariant features from speaker utterances.
\end{abstract}


% SAEP uses more than $98\%$, $94\%$, $95\%$, and $73\%$ less parameters compared to VGG-M, ResNet-34, ResNet-50, and x-vector, respectively. 
% (e.g. $94\%$ less compared to ResNet-34, and $73\%$ less parameters compared to x-vector). 

\section{Introduction}
\label{sec:intro}

Recently, there have been several attempts to apply Deep Learning (DL) in order to build speaker embeddings \cite{variani2014deep,safari2016features,snyder2017deep,bhattacharya2017deep,snyder2018x}. Speaker embedding is often referred to a single low dimensional vector representation of the speaker characteristics from a speech signal. It is extracted using a \gls{nn} model. There are different architectures proposed to extract these speaker representations either in a supervised (e.g., \cite{variani2014deep,snyder2017deep}) or in an unsupervised way (e.g., \cite{safari2016features,GHAHABI201816}). In the supervised methods, usually a deep architecture is trained for classification purposes using speaker labels on the development set. Then in the testing phase, the output layer is discarded, the feature vectors of an unknown speaker are given through the network, and the pooled representation from one or more hidden layers are considered as the speaker embedding \cite{snyder2017deep,snyder2018x}. 

On the other hand, attention mechanisms are one of the main reasons of the success of sequence-to-sequence (seq2seq) models in tasks like \gls{nmt} or \gls{asr} \cite{bahdanau2014neural,vaswani2017attention,chan2015listen}. In seq2seq models, these algorithms are applied over the encoded sequence in order to help the decoder to decide which region of the sequence must be either translated or recognized. Self-attention, as a specific type of attention, is the process of applying the attention mechanism to every position of the input itself.
% Unlike normal attention, people in \cite{} proposed a self-attention mechanism. In this type of attention instead of applying to the input sequence, as a specific type of attention, is the process of applying the attention mechanism to every position of the input itself. 
Self-attention has shown promising results in a variety of \gls{nlp} tasks, such as \gls{nmt} \cite{vaswani2017attention}, semantic role labelling \cite{strubell2018linguistically}, and language representations \cite{devlin2018bert}. The popularity of the self-attention as employed in networks such as \textit{Transformer} \cite{vaswani2017attention}, lies in its high parallelization capabilities in computation, and its flexibility in modeling dependencies regardless of distance by explicitly attending to the whole signal. Multi-head attention is a newly emerging attention mechanism, which is originally proposed in \cite{vaswani2017attention} for a \textit{Transformer} architecture and appeared very effective in many seq2seq models such as \cite{vaswani2017attention,chiu2018state,dehghani2018universal}. People in \cite{zhu2018self,india2019self} employ multi-head attention mechanism in the pooling layer to aggregate the information over multiple input segments and output a single fixed-dimensional vector which is further used in the \gls{dnn} classifier.

For speaker recognition, attention mechanism has been studied mostly in the pooling layer as an alternative to statistical pooling. In \cite{bhattacharya2017deep, zhang2016end, chowdhury2017attention}, attention is applied over the hidden states of a \gls{rnn} in order to pool these states into speaker embeddings for text-dependent speaker verification. In \cite{cai2018exploring}, a unified framework is introduced for both speaker and language recognition using attention. 
% Several works have presented improved self-attention based pooling layers. 
In \cite{zhu2018self}, a multi attention mechanism based on \cite{lin2017structured} is proposed to use more than one attention model to capture different kinds of information from the encoder features. Other works like \cite{india2019self} have explored the use of multi-head attention for the pooling layer.
% , which is also a newly emerging attention mechanism, and is originally proposed in \cite{vaswani2017attention} for the \textit{Transformer} architecture. 

% Nowadays, end-user applications running on mobile devices become more and more popular. The computational limitations of these mobile devices oblige the applications to use models which comply with certain constraints such as energy consumption, memory and storage size. However, deep learning approaches, usually are energy expensive, demanding storage, processing power, and memory.
% Most of them need dedicated hardware, like Graphics Processing Units (GPUs), to have a reasonable inference time. 
% These obstacles motivate researchers for the design of more efficient deep models. 
Computational limitations of the mobile devices oblige the end-user applications to use models which comply with certain constraints such as energy consumption, memory and storage size. However, these constraints are not usually met by deep learning approaches. These obstacles motivate researchers for the design of more efficient deep models. \textit{MobileNets} were proposed in \cite{howard2017mobilenets,sandler2018mobilenetv2} and showed promising results for image classification. They are built based on depth-wise separable convolutions \cite{sifre2014rigid} and further used in \textit{Inception} models \cite{ioffe2015batch} to reduce the computation in the first few layers. Architectures based on factorized convolution were also found to be successful in \cite{jin2014flattened} and \cite{wang2017factorized} for image classification tasks.
% \cite{wang2017factorized} introduces a similar factorized convolution as well as the use of topological connections. 
% Subsequently, the Xception network \cite{chollet2017xception} demonstrated how to scale up depthwise separable filters to outperform Inception V3 networks.
In another attempt, a bottleneck strategy was utilized in \cite{iandola2016squeezenet} to design a very small network. 
% Another small network is Squeezenet \cite{iandola2016squeezenet} which uses a bottleneck approach to design a very small network. 
There are other reduced computation networks such as structured transform networks \cite{sindhwani2015structured} and deep fried \glspl{cnn} \cite{yang2015deep}. 
% A different approach for obtaining small networks is shrinking, factorizing or compressing pretrained networks. Compression based on product quantization \cite{wu2016quantized}, hashing \cite{chen2015compressing}, and pruning, vector quantization and Huffman coding
% \cite{han2015deep} have been proposed in the literature. Additionally various factorizations have been proposed to speed up pretrained networks \cite{jaderberg2014speeding,lebedev2014speeding}. Another method for training
% small networks is distillation \cite{hinton2015distilling} which uses a larger network to teach a smaller network. Another emerging approach is low bit networks \cite{courbariaux2014training,rastegari2016xnor,hubara2017quantized}. % TODO: write about these models in ASR and also speaker recognition

In this paper, we present an end-to-end speaker embedding extractor for speaker verification inspired by self-attention networks. It is shown that self-attention mechanisms are able to capture time-invariant features from one's speech, which can result in a discriminative representation of the speaker. The end-to-end architecture comprises an encoder, a pooling layer, and a \gls{dnn} classifier. The main contribution of this work is to use an encoder and pooling layer solely relied on self-attention and feed-forward networks to create discriminative speaker embeddings. In order to show the effectiveness of the proposed approach, these embeddings are evaluated on \textit{Voxceleb1} \cite{Nagrani17}, \textit{Voxceleb2} \cite{Chung18b}, and \textit{VoxCeleb1-E} \cite{Chung18b} protocols. The preliminary results obtained on this attention-based system show superior performance compared to the x-vector baseline. Its performance is also competitive to some other \gls{cnn}-based benchmarks while having much less parameters. More precisely, our proposed system employs more than $98\%$, $94\%$, $95\%$, and $73\%$ less parameters compared to VGG-M, ResNet-34, ResNet-50, and x-vector, respectively. This is a  substantial reduction in model size which makes it a great candidate for resource-constrained devices. To the best of our knowledge, there is only one study to address concise deep models for speaker identification on mobile devices \cite{nunes2020mobilenet1d}, which is evaluated on TIMIT \cite{garofolo1993darpa} and MIT \cite{woo2006mobile} datasets.
% , and our contribution is the first paper for text-independent speaker verification to address the efficiency of deep models in terms of number of parameters. is more efficient in extracting time-invariant features from speaker utterances, while  making it a great candidate for resource-constraints devices

\section{Proposed Architecture}
\label{sec:self-attention nets}

\begin{figure}[t!]
\centering
\includegraphics[scale=.17]{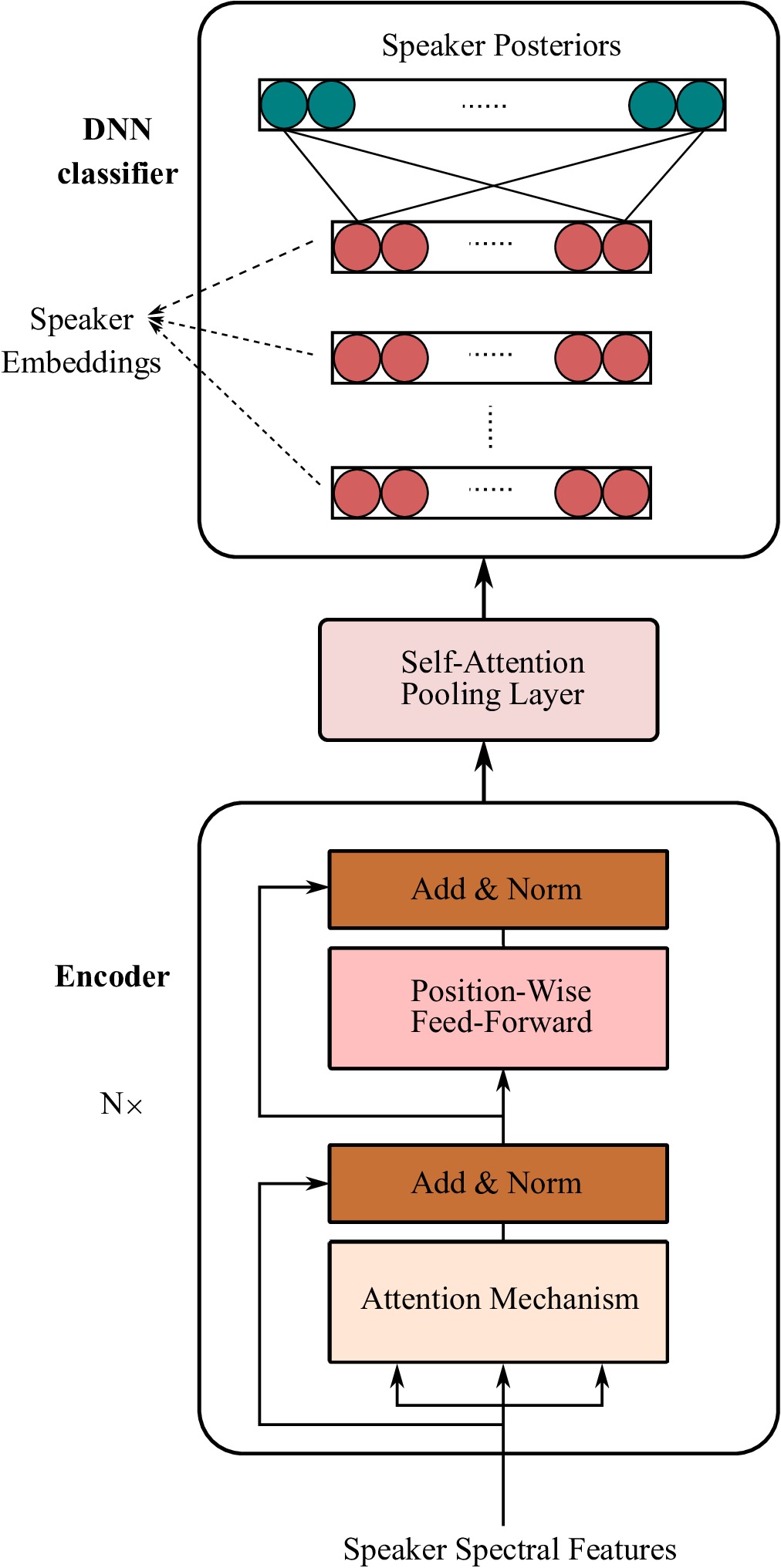}
\caption{The diagram of the proposed architecture. Encoder comprises a stack of N layers each of which is a series of attention mechanisms followed by position-wise feed-forward networks. A residual connection and layer normalization are applied to each of the sub-layers.}
\label{fig:System-Diagram}
\vspace{-.3cm}
\end{figure}

The system presented in this work can be divided into three main stages, namely an encoder, a pooling layer, and a \gls{dnn} classifier (Figure \ref{fig:System-Diagram}). 
The encoder is a stack of N identical blocks each of which has two sub-layers. The first sub-layer is a single-head self-attention mechanism, and the second is a %simple, --> do we need to say it's simple ?
position-wise feed-forward network. A residual connection is employed around each of the two sub-layers, and it is followed by a layer normalization \cite{ba2016layer}. The output of the encoder network is a sequence of feature vectors. The pooling mechanism is then used to transform these encoded features into an overall speaker representation. Finally, the third stage of the network is a \gls{dnn} classifier, which outputs the speaker posteriors. The activations of one or more of these hidden layers can be considered as the speaker embeddings. These embeddings can be further used in speaker verification using cosine scoring or more sophisticated back-ends.

\subsection{Self-Attention Encoder}
\label{ssec:encoder}

The input to each of the encoder blocks is passed through a self-attention network to produce representations by applying attention to every position of the input sequence. This is done using a set of linear projections.
% , which can in practice be implemented by means of highly optimized matrix multiplication. 
Consider an input sequence $X=[x_1, x_2,\ldots,x_T]^{tr}$ of length $T$, with $x_t\in \mathbb{R}^{d_{m}} $, and a set of trainable parameters $\{W_{Q},W_{K}\}\in \mathbb{R}^{d_{m}\times d_{k}},W_{V}\in \mathbb{R}^{d_{m}\times d_{v}} $, the model transforms the input into namely queries $Q \in \mathbb{R}^{T\times d_{k}}$, keys $K\in \mathbb{R}^{T\times d_{k}}$, and values $V\in \mathbb{R}^{T\times d_{v}}$:

\begin{equation}
\centering
\begin{split}
Q = XW_{Q}, \ K = XW_{K}, \ V = XW_{V}
% K = XW_{K} \\
% V = XW_{V}  
\end{split}
\end{equation}

The output for each time instance $o_{t}$ of the attention layer is computed as:
% \vspace{-.1cm}
\begin{equation}
\centering
o_{t} = Attn(q_{t},K)V
\vspace{+.1cm}
\end{equation}

\noindent where $Attn(\cdot)$ is an attention function that retrieves the keys $K$ with the query $q_{t}$. There are two most commonly used attention functions namely additive attention \cite{bahdanau2014neural}, and dot-product attention \cite{luong2015effective}. Here the attention is a scaled version of the dot-product attention mechanism, which is originally proposed in \cite{vaswani2017attention}. It is much faster and more space-efficient in practice, compared to the additive attention. Once the dimension $d_{k}$ grows, the additive attention outperforms the dot-product attention \cite{luong2015effective}. Therefore the scaling factor comes in handy to fill the performance gap between the two while keeping its advantages. The final output representation $O$ of the attention mechanism is formulated as:

\begin{equation}
\centering
O = Softmax(\frac{QK^{T}}{\sqrt{d_{k}}})V
\end{equation}

The output of the attention mechanism is sent to the other sublayer, which is a position-wise feed-forward network. It is applied to each position separately and identically. This consists of two linear transformations with a ReLU activation in between:
\vspace{-.1cm}
\begin{equation}
\centering
FFN(h) = max(0, hW_{1} + b_{1})W_{2} + b_{2}
\vspace{-.15cm}
\end{equation}

\noindent where $h$ is the input and $FFN(h)$ is the output of the feed-forward network. These linear transformations are the same across different positions, however, they differ from layer to layer. $W_{1} \in \mathbb{R}^{d_{m}\times d_{ff}}$ and $W_{2}\in \mathbb{R}^{d_{ff}\times d_{m}}$ can be considered as two convolutions with kernel size of $1$, while $d_{ff}$ is the feed-forward dimension. 

\subsection{Self-Attention Pooling Layer}
\label{ssec:pooling}

%Miquel Version
Self-attention based encoder outputs a sequence of short-term speaker representations. Given these frame-level features, it is needed to convert them into an utterance level vector. In this work, we will use a self-attention pooling layer. This layer is an additive attention based mechanism, which computes the compatibility function using a feed-forward network with a single hidden layer. Given the sequence of encoded features $H=[h_1, h_2,\ldots,h_T]^{tr} \in \mathbb{R}^{T\times d_{m}}$, we compute the segment-level representation $C$ as:

\begin{equation}
\centering
C = Softmax(W_cH^{T})H
\end{equation}

\noindent where $W_c \in \mathbb{R}^{d_{m}}$ is a trainable parameter. This self attention can be also understood as a dot product attention where the keys and the values correspond to the same representation and the query is only a trainable parameter. Therefore $C$ is a weighted average of the encoder sequence of features. These weights  correspond to the alignment learned by the self-attention mechanism. This kind of attention pooling is different from the ones proposed in \cite{zhu2018self,okabe2018attentive}, where attention is applied to extract statistics necessary for statistical pooling. However, in this work, we replace the whole statistical pooling with a simple additive attention mechanism with less computational cost. 

\subsection{Speaker Embeddings}
\label{ssec:speaker embeddings}
The fixed length vector obtained from the pooling layer is fed into a \gls{dnn} classifier to output the speaker posteriors. This \gls{dnn} classifier is based on three fully connected layers and a softmax layer to compute a multi-class cross entropy objective for training purposes. 

Once the network is trained, embeddings are extracted from one of the \gls{dnn} layers after the non-linearity. Instead of using the previous layer to the softmax as the speaker embedding, we use the second previous layer like in \cite{zeinali2019improve}. This is done because it has been shown by \cite{snyder2017deep} that this layer generalizes better, so it contains speaker discriminative information less adapted to the training data.

\section{Experimental Setup}
\label{sec:experimental setup}
% \vspace{-.25cm}
The performance of the proposed system is assessed using both \textit{VoxCeleb1} \cite{Nagrani17}, and \textit{VoxCeleb2} \cite{Chung18b} datasets with three different protocols. \textit{VoxCeleb1} development set is used for training in the first set of experiments (referred to as Vox1 hereafter). It contains over $147,935$ utterances from $1,211$ speakers. The trained models are evaluated on the original \textit{VoxCeleb1} test set, which contains $18,860$ verification pairs for each clients and impostors test. For the second set of experiments (referred to as Vox2 hereafter), \textit{Voxceleb2} develpoment set is used for training. It comprises $1,092,009$ utterances among $5,994$ speakers. The \textit{VoxCeleb1} test set is used for the assessment. In the third set of experiments (referred to as Vox1-E hereafter), we use \textit{Voxceleb2} development set for training, and \textit{VoxCeleb1-E} for the test. \textit{VoxCeleb1-E} consists of $581,480$ random pairs sampled from the entire \textit{VoxCeleb1} dataset (development$+$test), covering $1,251$ speakers.

For all the systems that we have trained, \textit{librosa} toolkit \cite{brian_mcfee_2019_2564164} is employed to extract $30$-dimensional Mel-Frequency Cepstral Coefficients (MFCCs) with the standard $25ms$ window size and $10ms$ shift to represent the speech signal. The deltas and double deltas together are added to produce a $90$ dimensional input for the networks. No data or test-time augmentation have been used during training or test. All features are subject to Cepstral Mean Variance Normalization (CMVN) prior to training. All models are then trained on chunks of speech features with $300$ frames. During the test, embeddings are extracted from the whole speech signal. ReLU is used as the non-linearity for all the systems. Adam optimizer has been used to train all the models with standard values as proposed in \cite{kingma2014adam} and learning rate of 1e-4.

The architecture of x-vector baseline is as proposed in \cite{snyder2018x}. Speaker embeddings are extracted from the affine component of the first layer after pooling. No dropout is used as suggested in \cite{zeinali2019improve} since it does not improve the results for the verification task. This statement is also confirmed by our experiments. All \gls{tdnn} layers are followed by non-linearity and Batch Normalization \cite{ioffe2015batch}. For this baseline architecture we have also considered \gls{plda} back-end. 
% For this baseline architecture we will consider two different back-ends. The first back-end is cosine distance and the second one is more a sophisticated combination of LDA and PLDA. 
Given the speaker embeddings, these are centered, then subject to Linear Discriminant Analysis (LDA) dimension reduction to $200$ and length-normalized. Additionally, we have trained a PLDA to score the trials of these post-processesed represenations for the speaker verification task.  

For the SAEP configuration, we use a stack of $N=2$ identical layers with $d_{k}=d_{v}=512$, and position-wise feed-forward dimension of $d_{ff} = 2048$. For the encoder network a dropout of $0.1$ is considered and for the rest of the network dropout is fixed to $0.2$. The dimension of the first dense layer is equal to $90$ and all other dense layers are equal to embedding dimension which is $400$. The size of $400$ for the embedding is chosen just to be similar as the i-vector system. No experiments have been conducted to explore the ideal size for the embedding. For SAEP, we have also tried Additive Margin Softmax (AMSoftmax) as originally proposed in \cite{wang2018additive} with scaling factor of $30$ and $0.4$ margin as the hyperparameters.

\begin{table*}[!t]
     \caption{Evaluation results on VoxCeleb data sets. Cosine distance is employed for scoring  unless  otherwise  stated. \textbf{Vox1}: train on Vox1-dev \& test on Vox1-test. \textbf{Vox2}: train on Vox2-dev test on Vox1-test. \textbf{Vox1-E}: train on Vox2-dev test on extended VoxCeleb1 test set. The result reported for ResNet-50 on Vox1-E, takes advantage of the test time augmentation \cite{Chung18b}. 
    %  The cosine distance is employed for scoring unless otherwise stated.
     }
     \vspace{.25cm}
% The result reported for ResNet-50 on Vox1-E, takes advantage of the test time augmentation by sampling ten 3-second temporal crops from each test segment, compute the distances between the every possible pair of crops
% $(10 \times 10 = 100)$ from the two speech segments, and use the
% mean of the $100$ distances 
   \label{TAB:results}
    \centering
    \begin{tabular}{llccccc}
        % \toprule
        &\textbf{Method} & Input & Loss & Dim.& \# Params &   \textbf{EER\%} \\ \midrule
        \multirow{5}{*}{\begin{turn}{90}\hspace{-.75cm}Vox1\end{turn}} &i-vector/PLDA \cite{Nagrani17} & MFCC &  - &$400$ & - &  8.8 \\ \cline{2-7}
        &VGG-M \cite{Nagrani17} & Spectrogram &  Softmax & $1024$& $67$M &  $10.2$\\
        &VGG-M \cite{Nagrani17} & Spectrogram &  Softmax+Contrastive & $1024$ & $67$M &   $7.8$\\
        % &x-vector & MFCC &  Softmax & $512$ & $4.38$M &   $ 8.56 $ \\ \cline{2-7}
        &x-vector & MFCC & Softmax &$512$ & $4.38$M & $12.62$  \\
        &x-vector/LDA/PLDA & MFCC & Softmax &$512$ & $4.38$M & $7.83$ \\ \cline{2-7}
        % &x-vector/LDA/PLDA & MFCC & Statistical & Softmax & 512 & $4.38$M  &  $7.32$ \\ \cline{2-8}
         &SAEP & MFCC &  Softmax & $400$& $1.16$M  & $7.70$  \\ 
         &SAEP & MFCC &  AMSoftmax & $400$ & $1.16$M &    $7.13$\\ \midrule
        \multirow{5}{*}{\begin{turn}{90}\hspace{-.5cm}Vox2\end{turn}} &   ResNet-34 \cite{Chung18b}& Spectrogram &  Softmax+Contrastive & $512$ & $21.8$M$^{\ast}$ &   $5.04$\\
        &ResNet-50 \cite{Chung18b}& Spectrogram &  Softmax+Contrastive & $512$ & $25.6$M$^{\ast}$ &  $4.19$ \\ 
        % &x-vector & MFCC & Softmax &$512$ & $4.38$M &  $7.38$ \\ \cline{2-7}
        &x-vector & MFCC & Softmax &$512$ & $4.38$M &  $11.64$ \\
        &x-vector/LDA/PLDA & MFCC & Softmax &$512$ & $4.38$M & $7.87$ \\ \cline{2-7}
         &SAEP  & MFCC &  Softmax & $400$ & $1.16$M &  $6.32$  \\ 
         &SAEP  & MFCC &  AMSoftmax & $400$ & $1.16$M &  $5.44$\\ \midrule
        \multirow{5}{*}{\begin{turn}{90}Vox1-E\end{turn}} & ResNet-50 \cite{Chung18b}& Spectrogram &  Softmax+Contrastive & $512$ & $25.6$M$^{\ast}$ &  $4.42$ \\ 
        % &x-vector & MFCC &  Softmax &$512$ & $4.38$M &  $8.06$ \\ \cline{2-7}
        &x-vector & MFCC & Softmax &$512$ & $4.38$M &  $11.62$ \\
        &x-vector/LDA/PLDA & MFCC & Softmax &$512$ & $4.38$M & $8.18$  \\ \cline{2-7}
        % &x-vector/LDA/PLDA & MFCC & Statistical & Softmax &$512$ & $4.38$M & $6.81$ \\ \cline{2-8}
         &SAEP  & MFCC &  Softmax & $400$ & $1.16$M & $6.94$  \\ 
         &SAEP  & MFCC &  AMSoftmax & $400$ & $1.16$M &  $5.62$ \\
         \midrule 
    \end{tabular}\\
    \raggedright \hspace{2cm} $^{\ast}$Minimum estimation, actual number not provided by the reference.
    \vspace{-.2cm}
\end{table*}

\section{Results}
\label{sec:results}

Results for text-independent speaker verification task are summarized in Table \ref{TAB:results} in three different sections for three different protocols. The cosine distance is employed for scoring unless otherwise stated. The performance of different systems have been evaluated using \gls{eer}. 
% In addition to these experiments we have also tried our pooling for the x-vector, however, no improvement has been seen. It seems that this kind of pooling which solely relied on attention cannot improve x-vector performance.
% The results for our x-vector baseline is consistent with the ones reported in \cite{shon2018frame} for non-augmented data on Vox1 protocol. 
% X-vector models with our pooling and with AMS have also been considered, but we have not been able to train them in the proposed protocols.
The proposed architecture is compared with i-vector, and also x-vector system, which is nowadays one of the most popular end-to-end speaker embedding extractors. Additionally, in order to analyse the trade-off between performance and network size, we have also included results reported by some other approaches such as VGG-M \cite{Nagrani17}, ResNet-34, and ResNet-50 \cite{Chung18b}. These models benefit from much larger amount of parameters.

For the Vox1 protocol we have compared SAEP with x-vector and the VGG based architecture proposed in \cite{Nagrani17}. It is the smallest protocol in terms of development data. 
% The proposed method shows a similar performance to both x-vector with LDA/PLDA back-end and VGG. 
SAEP with vanilla softmax 
% only 
shows a small relative improvement in terms of EER compared to x-vector with LDA/PLDA, and 
% only a $1.28\%$ 
% compared to 
VGG-M. 
However, using AMSoftmax this improvement increases to a $8.93\%$ in comparison with x-vector/LDA/PLDA and $7.99\%$ compared to VGG-M.

For the Vox2 and Vox1-E protocol we have added ResNet-34 and ResNet-50 based models, which have shown recently very competitive results for the speaker verification task \cite{Chung18b}. %I need here more cites. ResNet 34 IS2019 papers 
% In contrast to Vox1 protocol results, 
Here SAEP outperforms x-vector by a larger margin. The presented approach with softmax loss shows almost $20\%$ relative improvement in terms of \gls{eer} compared to x-vector with LDA/PLDA back-end in the Vox2 protocol and more than $15\%$ in the Vox1-E protocol. Compared to the ResNet standard architectures, SAEP is able to perform similar to ResNet-34 with more than $94\%$ less parameters. 
ResNet-34 and ResNet-50 have better results than our method and also x-vector, mainly due to the use of much larger number of parameters and more sophisticated training strategies such as contrastive loss. Furthermore, ResNet-50 takes advantage of test-time augmentation \cite{Chung18b} for Vox1-E protocol. 
% By looking at the table, it is realized that SAEP uses more than $98\%$, $94\%$, $95\%$, and $73\%$ less parameters compared to VGG-M, ResNet-34, ResNet-50, and x-vector, respectively. 

\begin{figure}[h!]
\centering
\includegraphics[scale=.7]{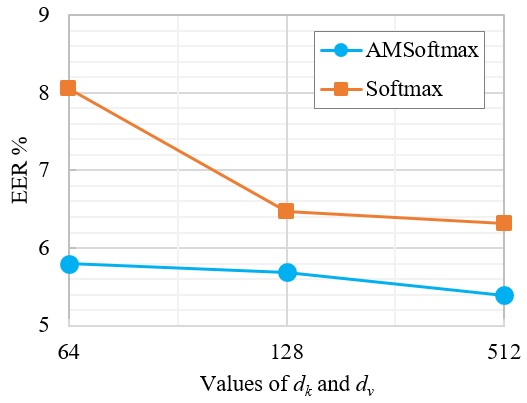}
\caption{Assessment of the impact of the key and value dimensions ($d_{k}=d_{v}$) on the performance of speaker embeddings in terms of EER\% for SAEP architecture. The results reported for both Softmax and AMSoftmax training losses on Vox2 protocol.}
\label{fig:dim-compare}
\vspace{-.2cm}
\end{figure}

In Figure \ref{fig:dim-compare} we have presented the effect of key and value dimensions of $d_{k}=d_{v}$ on the performance of our proposed system for both Softmax and AMSoftmax losses on Vox2 protocol. We have tried three different dimensions $64$, $128$, and $512$. It is notable to consider the total number of network parameters for each one of these systems. These dimensions correspond to $0.83$M, $0.88$M, and $1.16$M total number of model parameters, respectively. In addition to these experiments, in order to show the efficiency of our architecture in terms of number of model parameters, we tried an alternative configuration with $d_{ff}=1024$ and $d_{v}=d_{k}=64$. This configuration achieved $7.83\%$ EER on Vox2 protocol and Softmax loss, with only $0.45$M parameters. Compared to x-vector, which offers similar performance, this alternative approach  requires only almost one-tenth of the parameters required for the x-vector. This is a remarkable improvement, which justifies the superiority of our solution for resource-constraint devices, such as mobile phones. 

% highlight them in the abstract, in the conclusions, and in the results discussion. It is very difficult to catch this point while looking at a table with too many parameters

% This shows the huge difference between number of parameters of our model and other models presented in Table \ref{TAB:results} based on \gls{cnn} and \gls{tdnn}. 

% In comparison to x-vector/LDA/PLDA, which offers similar performance (i.e., 7.87% EER), our architecture requires only one-tenth of the parameters required for CNN and TDNN. This is a remarkable improvement, which justifies the superiority of our solution for resource-constraint devices, such as mobile phones. 

\section{Conclusions}
\label{sec:Conclusion and Future Work}
% \vspace{-.25cm}
We have presented a tandem encoder and pooling layer solely based on self-attention mechanism and position-wise feed-forward networks. They are used in an end-to-end embedding extractor for text-independent speaker verification. Preliminary results show that this fully attention based system is able to extract the time-invariant speaker features and produce discriminative representation for speakers with much fewer number of parameters compared to some other \gls{cnn} and \gls{tdnn}-based benchmarks. SAEP uses more than $98\%$, $94\%$, $95\%$, and $73\%$ less parameters compared to VGG-M, ResNet-34, ResNet-50, and x-vector, respectively. 
% (e.g. $94\%$ less compared to ResNet-34, and $73\%$ less parameters compared to x-vector). 
% we conclude that the fully attention based approach is efficient to extract time-invariant features. Considering the number of parameters involved for the SAEP, 
It reveals that SAEP is superior at capturing long-range dependencies compared to other models. One reason is that the attention mechanism looks into all time instances at once so it is more flexible to extract features,  which are located at longer distances from one another.
% we conclude that our fully attention based model is much more efficient in capturing speaker specific features. 
For the future work, we would like to study the effects of various data augmentation strategies since it is very effective on speaker embeddings such as x-vector. We also need to study the scale of the system with more encoding layers, larger dimensions, and the addition of an appropriate back-end.

\section{Acknowledgements}
This work was supported in part by the Spanish Project DeepVoice (TEC2015-69266-P).
\clearpage
\bibliographystyle{IEEEtran}
\bibliography{template}
\end{document}

%% file: glossaries.tex
\newacronym{dl}{DL}{Deep Learning}
\newacronym{dbn}{DBN}{Deep Belief Network}
\newacronym{dnn}{DNN}{Deep Neural Network}
\newacronym{rbm}{RBM}{Restricted Boltzmann Machine}
\newacronym{cnn}{CNN}{Convolutional Neural Network}
\newacronym{rnn}{RNN}{Recurrent Neural Network}
\newacronym{lstm}{LSTM}{Long Short-Term Memory}
\newacronym{dbm}{DBM}{Deep Boltzmann Machine}
\newacronym{cd}{CD}{Contrastive Divergence}
\newacronym{urbm}{URBM}{Universal RBM}
\newacronym{udbn}{UDBN}{Universal DBN}
\newacronym{relu}{ReLU}{Rectified Linear Unit}
\newacronym{vrelu}{VReLU}{Variable ReLU}
\newacronym{ubm}{UBM}{Universal Background Model}
\newacronym{gmm}{GMM}{Gaussian Mixture Model}
\newacronym{map}{MAP}{Maximum a Posteriori}
\newacronym{nist}{NIST}{National Institute of Standard and Technology}
\newacronym{sre}{SRE}{Speaker Recognition Evaluation}
\newacronym{nap}{NAP}{Nuisance Attribute Projection}
\newacronym{jfa}{JFA}{Joint Factor Analysis}
\newacronym{lid}{LID}{Language Identification}
\newacronym{lda}{LDA}{Linear Discriminant Analysis}
\newacronym{vad}{VAD}{Voice Activity Detection}
\newacronym{plda}{PLDA}{Probabilistic Linear Discriminant Analysis}
\newacronym{wccn}{WCCN}{Within-Class Covariance Normalization}
\newacronym{asr}{ASR}{Automatic Speech Recognition}
\newacronym{det}{DET}{Detection Error Tradeoff}
\newacronym{eer}{EER}{Equal Error Rate}
\newacronym{dcf}{DCF}{Detection Cost Function}
\newacronym{mindcf}{minDCF}{minimum DCF}
\newacronym{far}{FAR}{False Acceptance Rate}
\newacronym{frr}{FRR}{False Rejection Rate}
\newacronym{fbe}{FBE}{Filter Bank Energy}
\newacronym{ff}{FF}{Frequency Filtering}
\newacronym{hmm}{HMM}{Hidden Markov Model}
\newacronym{lpc}{LPC}{Linear Predictive Coefficient}
\newacronym{lpr}{LPR}{Log Posterior Ratio}
\newacronym{mfcc}{MFCC}{Mel-Frequency Cepstral Coefficient}
\newacronym{lfcc}{LFCC}{Linear Frequency Cepstral Coefficient}
\newacronym{nn}{NN}{Neural Network}
\newacronym{pca}{PCA}{Principal Component Analysis}
\newacronym{plp}{PLP}{Perceptual Linear Predictive}
\newacronym{sdc}{SDC}{Shifted Delta Cepstrum}
\newacronym{psd}{PSD}{Power Spectral Density}
\newacronym{rbf}{RBF}{Radial Basis Function}
\newacronym{dct}{DCT}{Discrete Cosine Transform}
\newacronym{snr}{SNR}{Signal-to-Noise Ratio}
\newacronym{svm}{SVM}{Support Vector Machine}
\newacronym{cmn}{CMN}{Cepstral Mean Normalization}
\newacronym{cmvn}{CMVN}{Cepstral Mean and Variance Normalization}
\newacronym{cdf}{CDF}{Cumulative Distribution Function}
\newacronym{pdf}{PDF}{Probability Density Function}
\newacronym{em}{EM}{Expectation-Maximization}
\newacronym{sv}{SV}{Speaker Verification}
\newacronym{sr}{SR}{Speaker Recognition}
\newacronym{an}{AN}{Adversarial Network}
\newacronym{en}{EN}{Encoding Network}
\newacronym{dn}{DN}{Discriminative Network}
\newacronym{mlp}{MLP}{Multilayer Perceptron}
\newacronym{gru}{GRU}{Gated Recurrent Unit}
\newacronym{tdnn}{TDNN}{Time Delay Neural Network}
\newacronym{nmt}{NMT}{Neural Machine Translation}
\newacronym{fc}{FC}{Fully Connected}
\newacronym{san}{SAN}{Self-Attention Network}
\newacronym{nlp}{NLP}{Natural Language Processing}
\newacronym{sabe}{SABE}{Self-Attention Based Encoder}